\documentclass[prb,aps,twocolumn]{revtex4}
\usepackage{graphicx}
\DeclareGraphicsExtensions{.jpg,.pdf,.png}
\usepackage{amsmath}
\begin{document}
\title{Crack (Non)Propagation in Amorphous Media}
\author{Tsviki Y. Hirsh}
\affiliation{Department of Physics, Bar-Ilan University, Ramat-Gan IL52900 Israel}
\author{David A. Kessler}
\affiliation{Department of Physics, Bar-Ilan University, Ramat-Gan IL52900 Israel}
\date{September 23, 2004}
\begin{abstract}
We study Mode I crack propagation in amorphous material via a molecular dynamics simulation of a binary alloy
with pairwise central-force interactions.  We find that when the system is subjected to 
constant displacement after introduction of a seed crack, the crack does not propagate.  This was
found to be true both for the modified Lennard-Jones potential of Falk and a shorter-range potential of our
own devising.  Crystalline samples prepared with these potentials exhibit normal brittle fracture, with
running cracks.  Only when subjected to a constantly increasing strain did the crack in the glasses lengthen.  
Even here,
once the strain stopped increasing, the crack tip stopped moving.  We examined the stress-strain curve for
our model materials, and found that they did not exhibit the saturated stress plateau characteristic of
ductile materials.  We attribute the failure of crack propagation in this system to the availability of
many meta-stable configurations available to the crack tip which serve to soak up the energy which would otherwise
drive the crack.  We conclude that this class of models is inadequate to describe crack propagation in brittle
glasses.
\end{abstract}
\maketitle
\section{Introduction}
The mechanism of crack propagation in amorphous materials, such as glasses, is still far from
understood. There is good reason to believe that it exhibits significantly differences from fracture
in crystalline solids.  The first difference is the existence of a velocity gap, which is a necessary feature
of a crystalline lattice system, but which has not been seen in experiments on amorphous systems.
The second is the behavior of high velocity cracks.  In crystalline lattices, above a certain velocity a Mode I
crack no longer propagates along the midline.  Rather, it chooses to bifurcate, and go off along some other
easy propagation direction.  This behavior has been seen both in experiments on crystalline silicon
and in simulations of two-dimension crystals.  Amorphous materials have no easy directions, and must
do something different.  Typically, the exhibit a form of "side-branching" behavior, in which daughter cracks
are emitted which extend for some short distance, while the main crack keeps on propagating.
A better theoretical grasp of fracture in amorphous materials is clearly desirable, in order to better 
characterize the novel features that such dynamics exhibits.
 
 Surprisingly, there have been relatively few simulational studies of
crack propagation in glasses, given that the vast majority of  experiments have been performed
using various brittle amorphous materials, with
only a small number  employing crystalline solids.  This is no doubt at least in part a reflection of the increased technical
difficulties in simulating glasses.  Even constructing a simulational sample of glass without external stresses
requires a delicate process of cooling, whereas an unstressed crystalline material is trivial to construct. 
A more subtle difficulty is connected to the nature of the length scales involved.  
Due to the singular nature of the continuum stress field around a crack tip, the fracture mechanism 
is exquisitely sensitive to the dynamics of bond-breaking at the atomic scale.  In a crystal, this is the only
scale, apart from the sample size itself, and thus it is relatively easy to achieve essentially macroscopic
results on systems only a few tens of lattice lengths wide.  The microscopic structure of glasses is much
more involved, with intermediate length scales that most also be resolved in order to achieve reliable
results.  

When surveying the existing simulational attempts to study crack propagation in glasses, one is immediately struck
by the paucity of results, compared with the corresponding situation in crystalline systems.  There has been,
to our knowledge, no result that reproduces the basic experimental setup, namely a material with a 
seed crack that is subjected to a fixed external displacement, and then slightly perturbed so as to 
start the crack running the length of the system.  Our basic point
in this paper is that this situation is due to the fact that, for the class of models of glasses that have been
examined to date, cracks {\em do not propagate}.  The cracks that have been generated have been a result
of systems which have been continually {\em pulled apart}, and not due to intrinsic crack propagation of the
sort that has been seen in experiment, as well as in simulations of crystalline materials.  The ductile/brittle
transition which has been discussed in the literature \cite{Falk} does not capture the distinction between freely running
cracks and cracks which exist only due to continual pulling, which is the relevant one for crack propagation studies.

In this paper, we will first describe our model and simulational methods. In Section III, we will describe the
stress-strain relationship we measure, and its implications.  In Section IV, we will examine the behavior of a
crack in our system.

\section{Model and Sample Preparation}

The basic simulational framework we employ in this work is a 2D Molecular-Dynamics (MD) simulation,
with a 2-body central force interaction between particles.  We use the standard technique of mixing two
sizes of particles in order to avoid crystallization.  Binary alloy systems of this type are known to exhibit a glass-transition
when cooled from an initial high-temperature liquid state.  This model is considered a good model for
metallic glasses.  Such metallic glasses exhibit  very plastic and ductile behavior, in contrast to the very 
brittle glass that is used for fracture experiments.  Such behavior was also seen by Falk and Langer\cite{FL} in his simulations
of the stress-strain curve for a binary alloy Lennard-Jones (LJ) glass.  Imposing a strain linearly
increasing with time, Falk and Langer
 found that after an initial linear elastic regime, the stress saturated at some finite value.
In other words, instead of breaking, the material began to undergo plastic flow.  Clearly such a model
is unsuitable for studying brittle fracture.  To overcome this problem, Falk~\cite{Falk} introduced a modified interaction potential,
which he termed a compressed Lennard-Jones (CLJ) potential.  This CLJ potential had a narrower potential well,
as well as a faster falloff at large distance as compared to the standard LJ potential.  Falk found that
with this new CLJ potential, the stress-strain relationship changed character, and now fell off beyond a critical
strain.  This is typically taken to be an indication of brittle behavior.  As we shall see, the CLJ model
nevertheless does not exhibit brittle fracture.  To see to what extent this is generally true,  we have employed
in our simulations both the CLJ potential as well as a second potential which we introduce here.  This new potential
was designed to be even more brittle than the CLJ potential.  Nevertheless, we found the qualitative fracture dynamics
of the two potentials to be identical.

To understand why the LJ potential generates such a ductile material, let us consider  the force law of the
LJ potential :

\begin{equation}
\boldsymbol{f}_{LJ}\left(r\right) = 48\frac{\varepsilon_{ij}}{\sigma_{ij}}\left[ \left( \frac {\sigma_{ij}} {r_{ij}}
\right) ^{13} - \frac 1 2 \left( \frac {\sigma_{ij}} {r_{ij}} \right) ^7 \right] \hat
{\boldsymbol{r}}_{ij},
\end{equation}

\begin{figure}
\includegraphics[width=.38\textwidth]{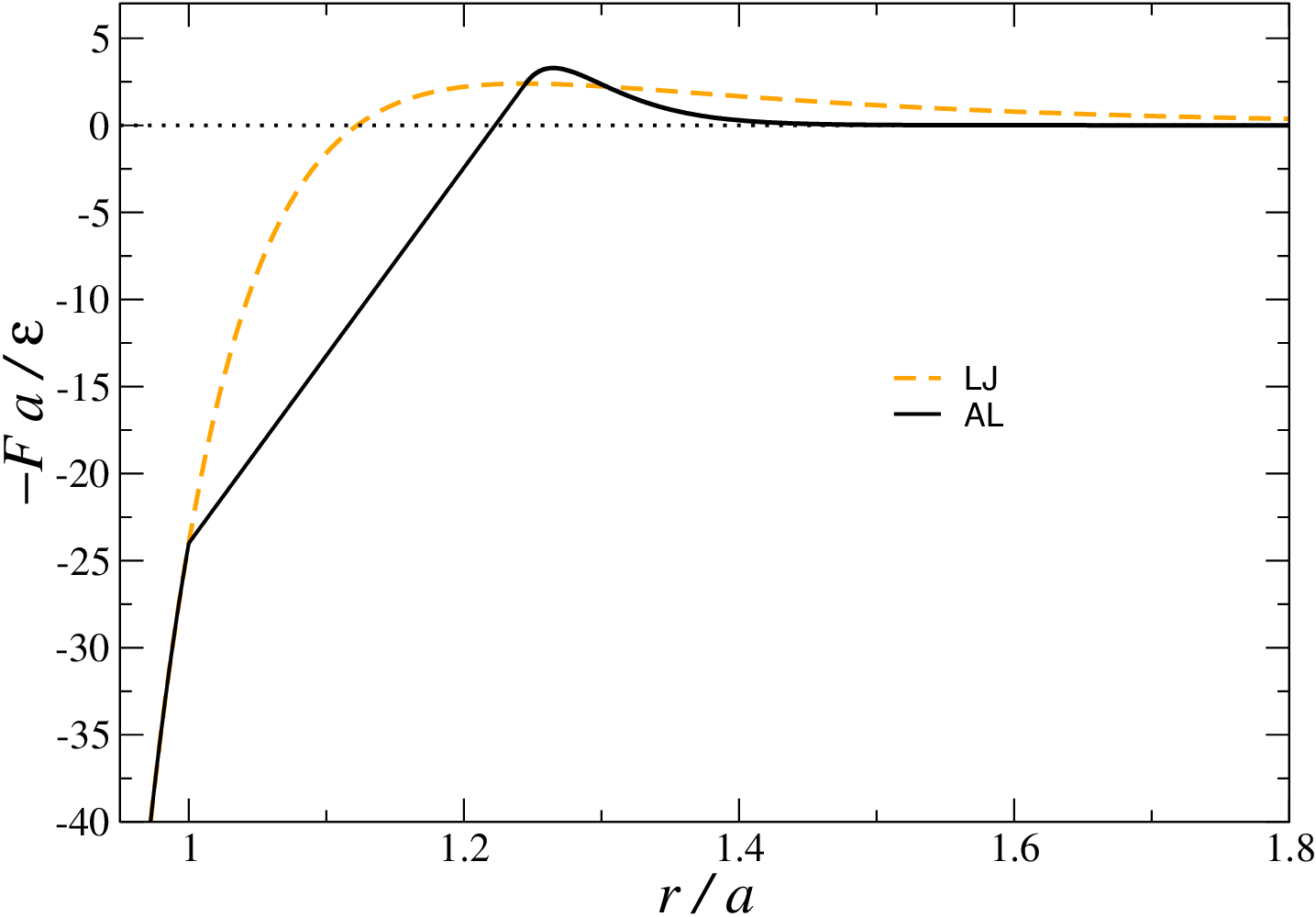}
\caption{The AL force law, compared to the standard LJ force}
\end{figure}

After the force maximum at $a_m \equiv \left(2\cdot2^{\frac{1}{6}}-1\right)\cdot{}a$,   there is a
very long attractive tail.  This character
of the LJ force law makes it very suitable for simulations of solidification. 
In order to have a force law that generates a brittle material, we have to shorten the attractive
tail. so that the material breaks almost immediately after a critical strain is applied.  The ideally
brittle force law is a 
"Piece-Wise Linear Force Law" namely a force law that shows a linear increase of force with
increasing strain until a critical strain, and a zero force afterwards break down. This force law is
the idealization of brittle behavior.
However we
cannot allow a too short tail, because it makes the glass solidification process very difficult.   Therefore,
we introduce a force law, which for convenient reference we call the AL (Almost Linear) force law,
 which is basically composed of a polynomial repulsive tail, as in  LJ,
 a linear force for small deformations from equilibrium and an exponential decay of
the force after the maximum force.  In detail

\begin{equation}
f_{AL}\left(r\right)=\left\{ \begin{array}{ll}
48\frac{\varepsilon_{ij}}{\sigma_{ij}}\left[ \left( \frac {\sigma_{ij}} {r} \right) ^{13} - \frac 1 2 \left( \frac
{\sigma_{ij}} {r} \right) ^7\right]~~~~~~~~r<a\\
\varepsilon_{ij}A+\varepsilon_{ij}B\cdot r~~~~~~~~~~~~~~~~~~~a<r<a_m\\
\varepsilon_{ij}C\left[ 1+D\left( r-a_m \right) \right] e^{-\beta\left( r-a_m\right)}~~~r>a_m\\
\end{array}\right .
\end{equation}

Where A,B,C,D depend on $\sigma_{ij}$ and are chosen so as to make the force continuous, and
$a_m = \left( 2\cdot 2^{\frac{1}{6}}-1\right)\cdot a$ is the location of the LJ maximum force.
The effective radius which corresponds to the potential minimum is $a_0 = 1.222a$   and the
potential well depth is $\varepsilon_0 = 2.66\varepsilon$.  For computational efficiency, we compute
the force via a cubic-spline interpolation, which ensures that the derivative of the force is continuous.

Our system consists of a binary mixture of particles,  80\% large particles (A-type) and
20\% small particles (B-type). Both types of particles have the same mass (set to be 1), and
different force parameters, as indicated in Table 1.

\begin{table}
\caption{Interaction parameters for binary mixture}
\begin{ruledtabular}
\begin{tabular}{ c c c}
 $\varepsilon$ & $\sigma$ & ~\\
 \hline
0.5 & 1 & AA \\
0.75 & 0.8 & AB \\
0.25 & 0.88 & BB \\
\end{tabular}
\end{ruledtabular}
\end{table}

The parameters were chosen in order to achieve  interaction properties similar to the LJ potential~\cite{Kob}.
Several other sets of parameters are employed in literature, designed to model
various specific glasses. However our interst is the fracture process and not in modeling some particular
real glass, and so any set of parameters which generates a glassy material is sufficient for our purposes. The MD simulations were preformed using a half step leap-frog algorithm, with a time step
of $\Delta t = 0.001$.  We start by randomly placing particles (A or B) on the simulation square, in the form of
a square lattice. Periodic boundary conditions were applied, so that the whole process is done
via constant density. The density is taken to be a bit larger than necessary, in order to avoid
creation of holes inside the glass.

We let the system to mix while holding the temperature in a very high value of
$5.1~\frac{\varepsilon_0}{k_B}$ throughout 20,000 time steps. The system temperature is controlled
by scaling of the particles velocities once every few time steps, according to the desired
temperature~\cite{Rapaport}.

After the system is mixed, we gradually reduce the temperature across the system by a temperature
decrease rate of $10^{-6} \frac{deg}{time step}$, to a temperature level of
$4.5~\frac{\varepsilon_0}{k_B}$. we let the system to relax by holding the temperature on a
constant value for 10,000 time steps. The process continues by gradually reducing the temperature
to lower temperature levels $\left(3.5,~2.5,~2,~1.8~...\right)$, while in every temperature level
we hold the temperature contant for 10,000 time steps.

This model is known to pass through a glass transition at about $0.3~\frac{\varepsilon_0}{k_B}$;
afterwards the particles configuration stays approximately fixed. After the system reaches a 
temperature of $0.05~\frac{\varepsilon_0}{k_B}$,  well under the glass transition, a small Kelvin type friction~\cite{Kessler_Kelvin},
$f_{ij}^K=\eta \left( v_i - v_j \right) $, with $\eta= 0.08$, is turned on in order to reduce
the temperature to zero. This process takes approximately another 100,000 time steps.
After the glass has been fully relaxed, we cut its edges from the periodic boundary conditions, and
again allow it to relax slowly to adjust to the change of pressure, giving us finally our rectangular piece of glass. 

Strain is imposed on our sample
 by displacing the last few rows of particles near the top and bottom edges to be displaced
according to the desired strain, and freezing any subsequent motion.  To impose an initial
seed crack, we solved for the continuum elastic solution of a seed crack of a given length in a rectangle
of the same aspect ratio as our sample.  This calculation was performed using a multigrid technique described in
Ref. \onlinecite{lefm,procac}.  We then interpolated this solution to the equilibrium locations of the particles
in our sample, and gave the particles the corresponding displacement.  We then relaxed our sample for some
time to allow it to return to equilibrium.

\section{Cracks} 

To study the dynamics of cracks in our system, the natural experiment to perform is
to seed an initial crack with some given external displacement, to perturb the crack tip
and analyze the crack motion.  This is exactly the experimental protocol used in the Fineberg
experiments.  However, when we attempted this using the CLJ potential, the crack refused to run.  No matter how
large the initial perturbation, or its precise form, the crack would almost instantly stop, with a reorganization of the particles in the vicinity of the crack tip.

To see if perhaps the potential was not "brittle" enough, we repeated the experiment using our new potential.
Again the crack tip refused to run.  As a test, we created a crystal with our new potential, and repeated the
experiment.  Now the crack had no problems running, producing results very similar to what is seen in
lattice simulations.

\begin{figure}
\includegraphics[width=.38\textwidth]{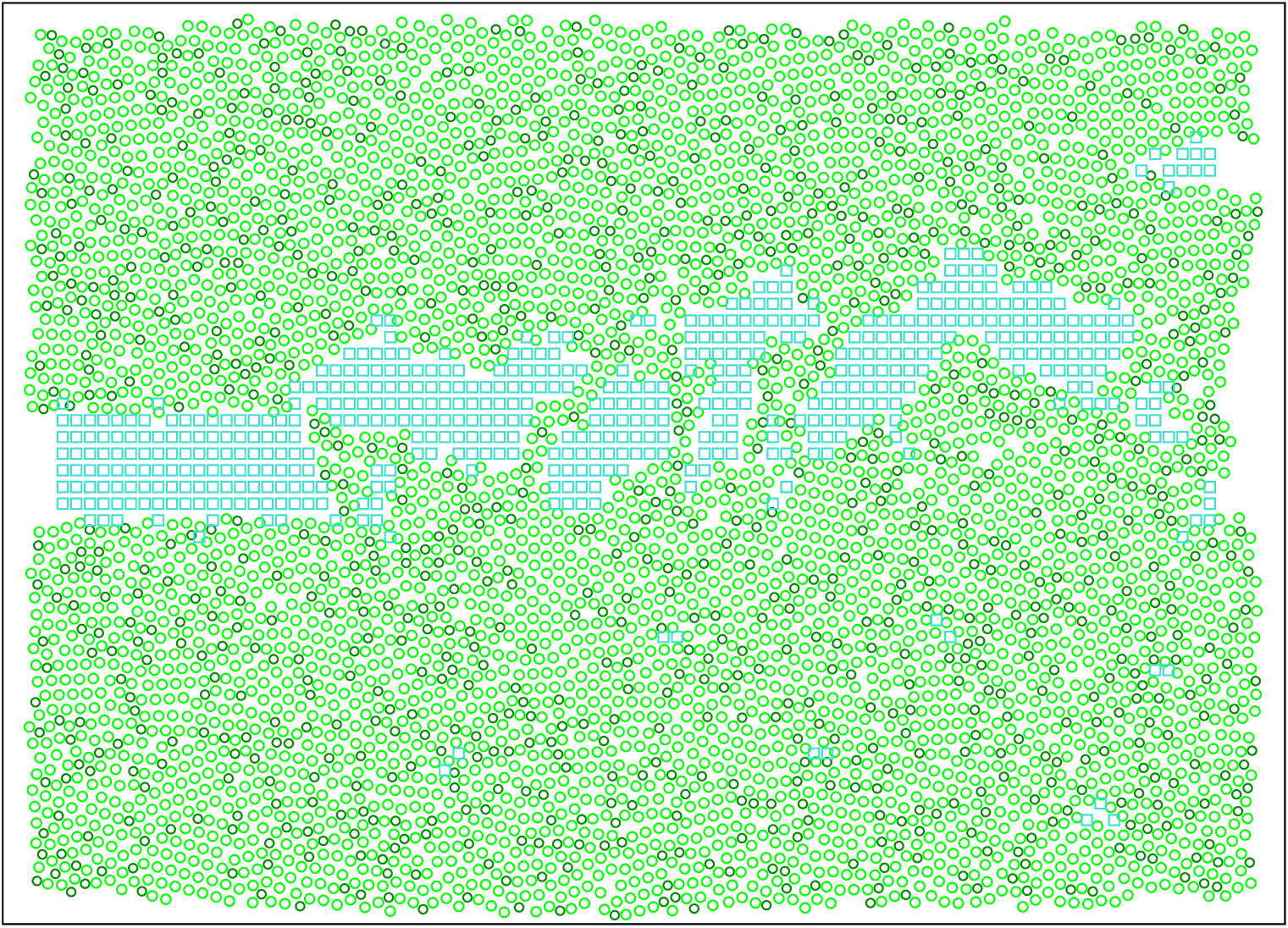}
\caption{Configuration of particles at end of run.  Strain rate=.  System size was $\times$.}
\label{latt}
\end{figure}

\begin{figure}
\includegraphics[width=.38\textwidth]{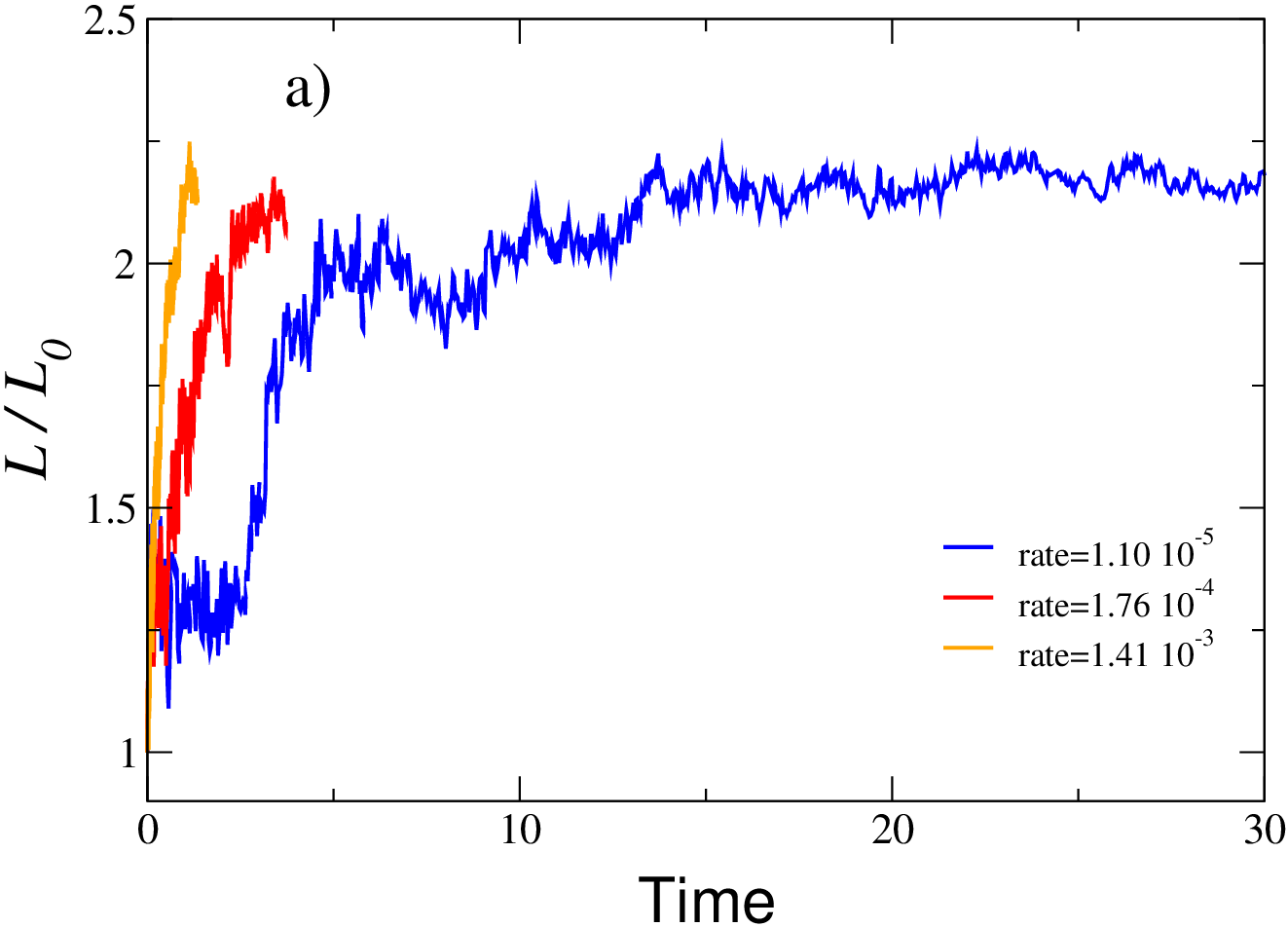}
\includegraphics[width=.38\textwidth]{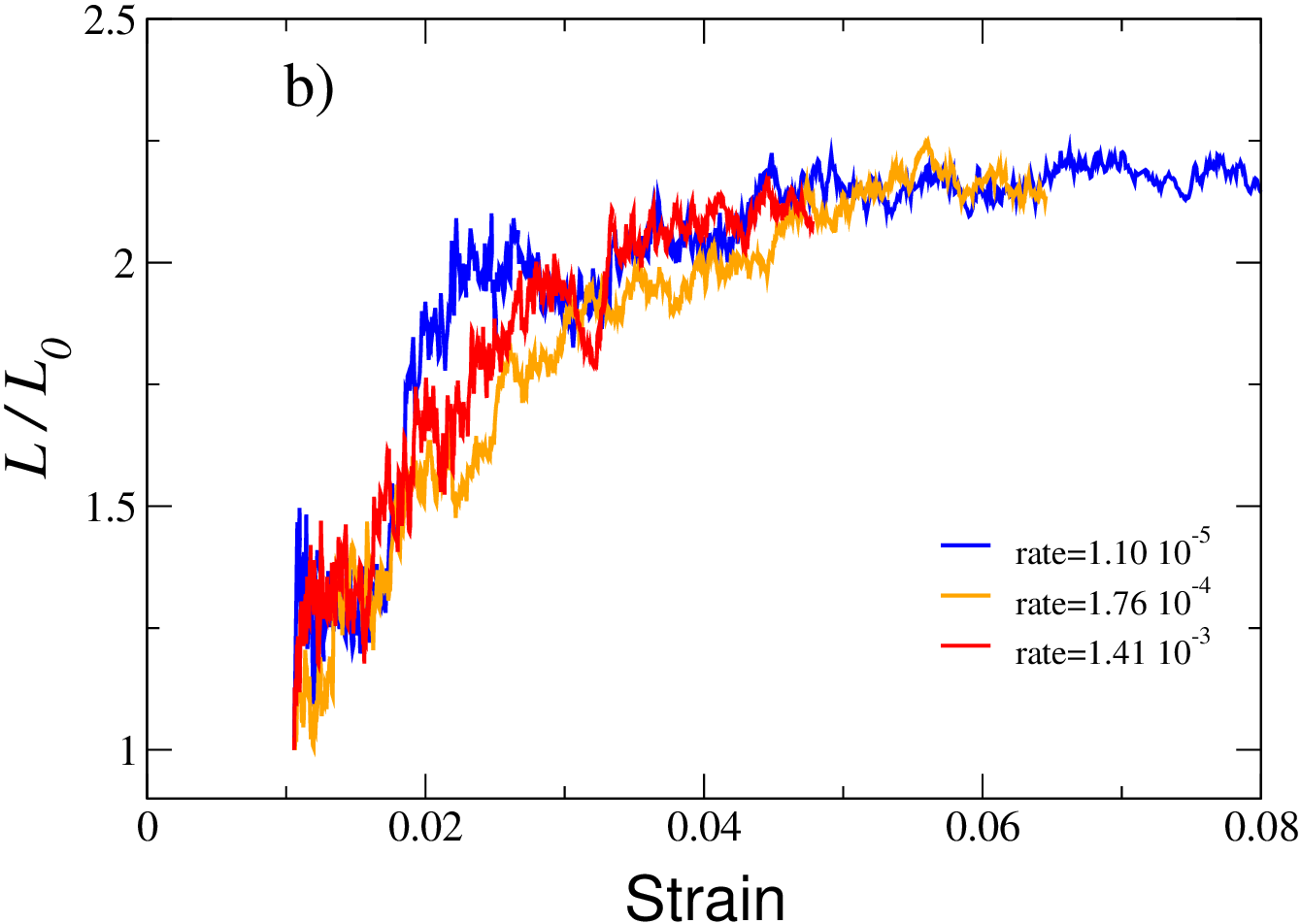}
\caption{a)Crack length versus time for three different strain rates. b)
Same date plotted as a function of strain.
The samples consisted of 6000 particles
with an aspect ratio of 1.6.  The length $L_0$ of the initial seed crack was 1/5 of the length of the sample.}
\label{many-strain}
\end{figure}

Given that Falk had produced "running" cracks with the CLJ potential, the difference must be attributed to
the different experimental protocol.  In Falk's work, instead of using a constant external displacement, he
employed a displacement growing linearly in time.  We then repeated our simulation, this time like Falk
using a displacement increasing linearly in time.  This did indeed produce "running" cracks.  
(See Fig. \ref{latt} for a picture of the crack at the end of the run.)  Clearly then,
the strain rate is an important parameter.  We ran simulations for a range of strain rates, producing the
results in Fig. \ref{many-strain}a.  The crack length indicated in the Figure was determined by partitioning the
sample into squares of size a lattice spacing, and counting the number of squares not containing
particles.  In this count, empty squares not touching another empty square were ignored, so as no
to include isolated voids in the measurement.  If instead of plotting crack length as a function of time,
one plots it versus the instantaneous external displacement, one sees in Fig. \ref{many-strain}b that all the curves
collapse.  Thus the crack length is a function of the external displacement only.  This is in complete contrast
to the standard dynamical fracture scenario, where the external displacement is {\em constant} in time, and
the crack length increases nonetheless, due to its own internal dynamics.  In these model glasses, however,
the crack dynamics is for some reason completely damped, and the crack responds only to continually increasing
external driving.

\begin{figure}
\includegraphics[width=.38\textwidth]{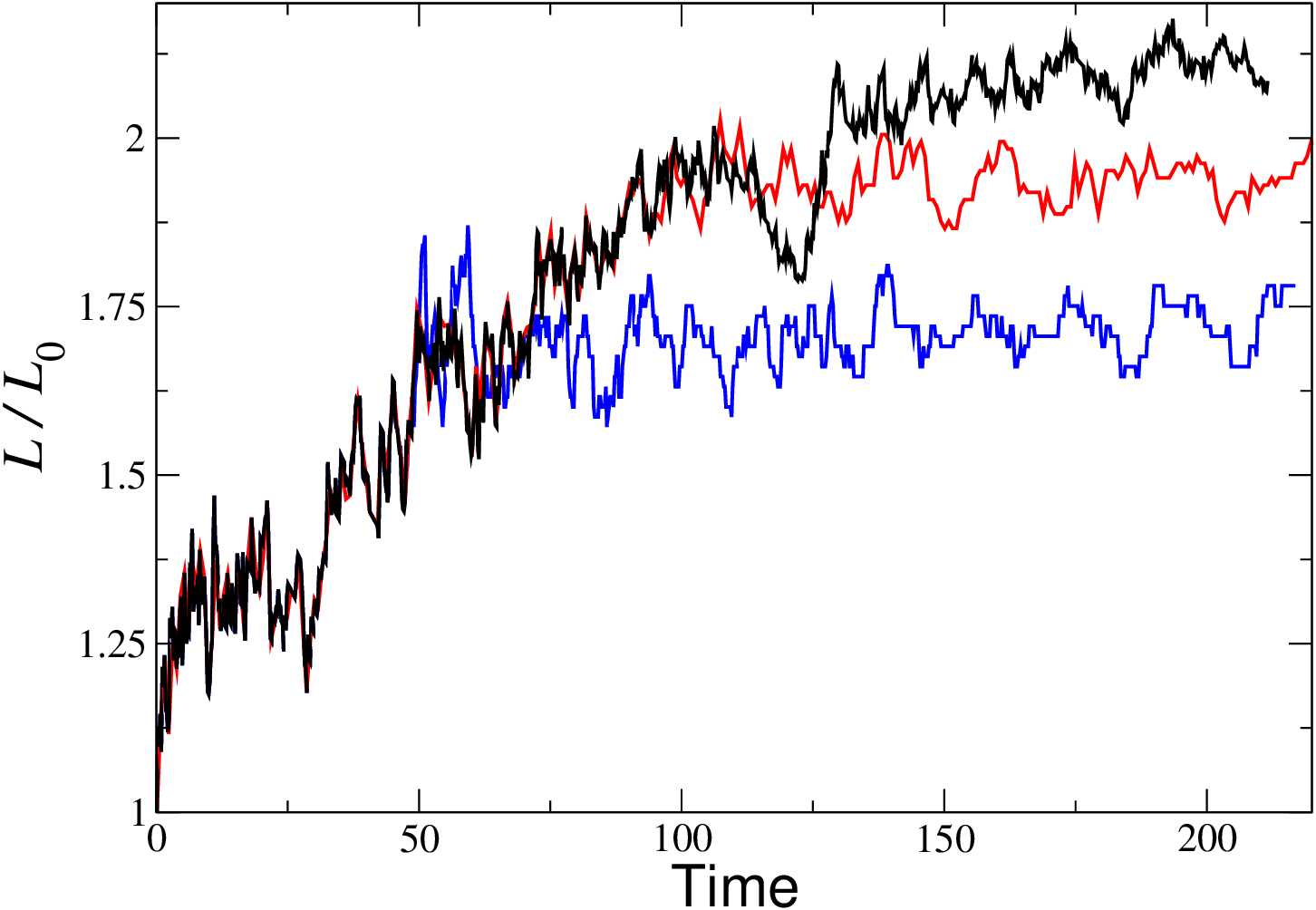}
\caption{Crack length versus time.  The top curve is the same data as in Fig. \protect{\ref{many-strain}}.
The lower curves portray simulations where the strain increase was turned off at various times.  The
immediate cessation of further fracture is evident.}
\label{stop-cracks}
\end{figure}

A striking way to verify this diagnosis is to turn off the strain increase in the middle of the run.  The results
of this are presented in Fig. \ref{stop-cracks}.  We see that as soon as the strain increase is turned off,
with the strain remaining constant afterwards, the crack stops dead in its tracks.  This is despite the fact
that the crack is propagating at a significant velocity prior to the turning off of the strain increase.

\section{Stress-Strain Relationship and Internal Damping}

We saw in the previous section that our model glasses did not support running cracks.  Thus must 
therefore possess
some sort of internal damping mechanism that is capable of soaking up the energy associated with the
stress release of the bulk as the crack lengthens.  In order to understand this better, we turn to a closer
examination of the stress-strain relationship.
The stress tensor element - $\sigma_{yy}$ is measured, as we increase the
strain -$\varepsilon_{yy}$ from zero, with a slow strain rate. The strains increased linearly with
time, $\Delta = b\cdot t$, where $b$ is the strain rate (dimensions of $\left[\frac{1}{time}\right]$
). The stress tensor is measured by using the virial expresion \cite{Rapaport}. 
 
\begin{equation}  
\sigma_{\alpha \beta} = \frac{NT_{\alpha \beta}}{V} + \frac{1}{2}\sum_{i<j}^N r_{ij}^\alpha\cdot
f_{ij}^{\beta}  
\end{equation} 
 
Where $\alpha , \beta \in xy$~$i,j~\in~1\ldots N$, $N$ is the number of particles, $V$ is the system
volume and $T_{\alpha\beta}=\frac{1}{N} \sum_i v_i^\alpha v_i^\beta$ indicates the system
temperature.

\begin{figure}
\includegraphics[width=.38\textwidth]{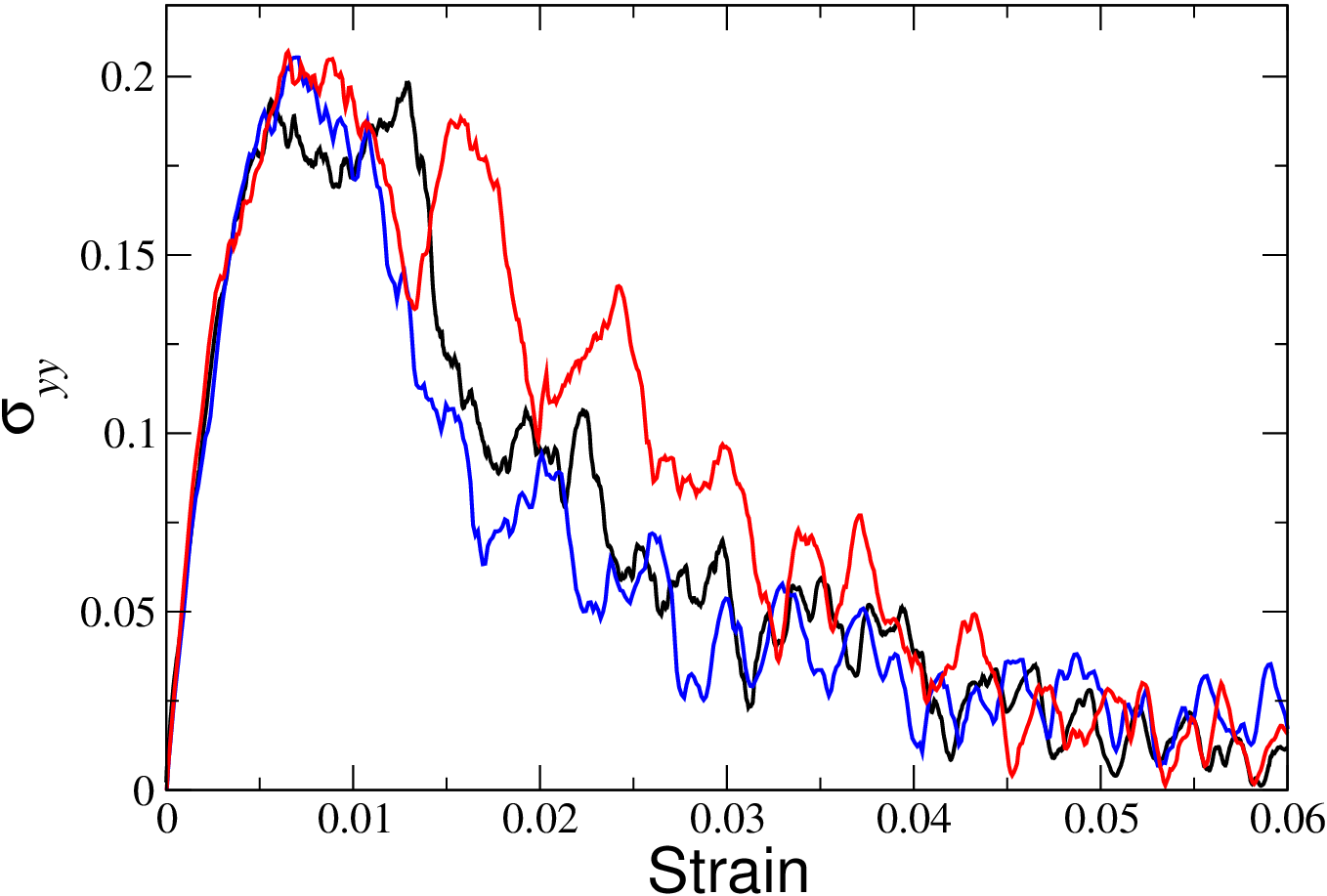}
\caption{Stress-Strain curves for three different samples of the same AL glass.  Strain rate was $1.76\cdot10^{-4}$.  Samples
were 6000 particles, with an aspect ratio of 1.6.}
\label{3samples}
\end{figure}

\begin{figure}
\includegraphics[width=.38\textwidth]{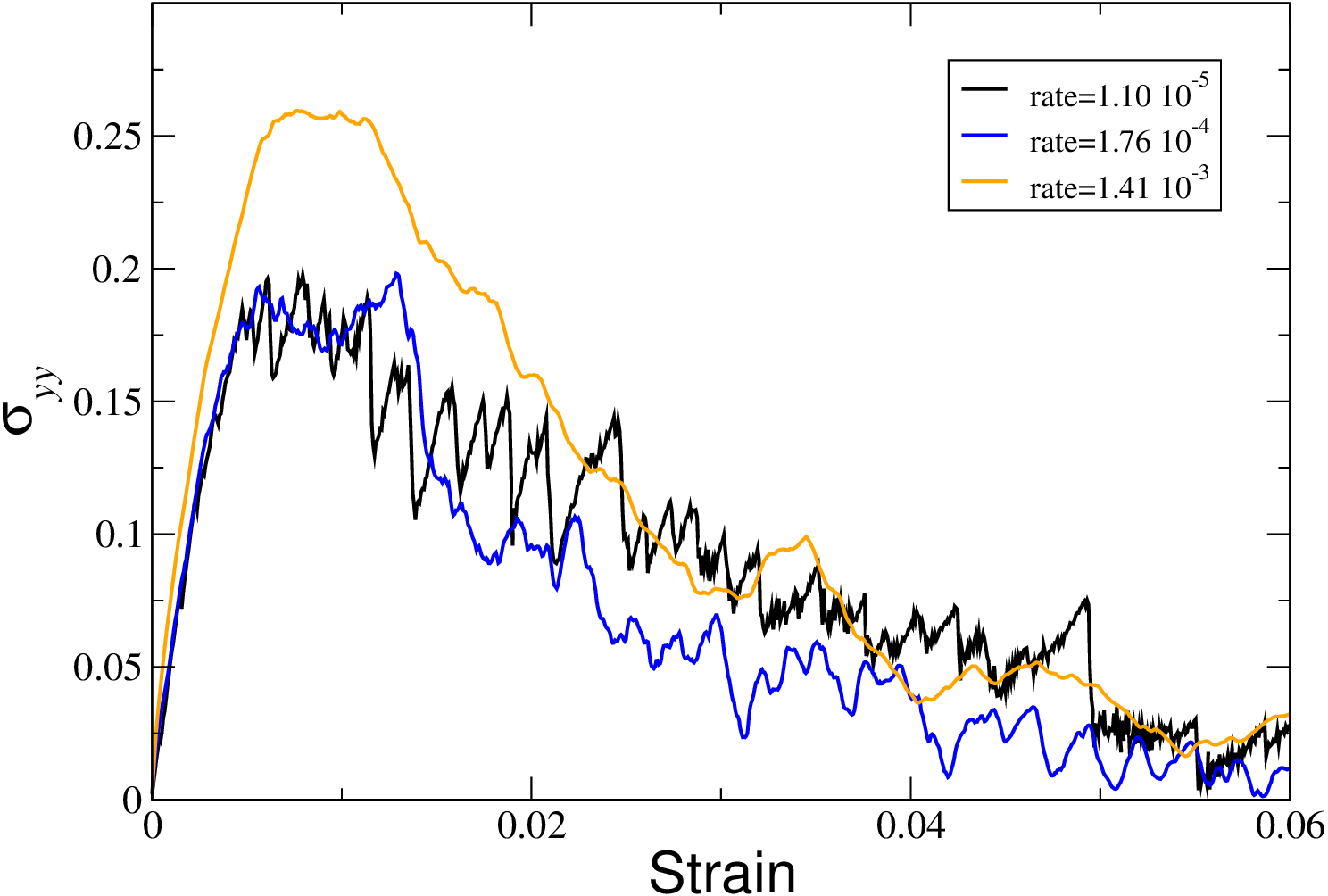}
\caption{Stress-strain curves for three different strain rates. The sample was as in Fig. \protect{\ref{3samples}}.}
\label{manyrates}
\end{figure}

Figure \ref{3samples} shows stress-strain curves for 3 different realizations of a glass formed using our AL force law. 
Each
simulation uses 6000 particles and a strain rate of $1.76\cdot10^{-4}$. In the region of small strains, the
material shows an elastic behavior, with reversible deformation. This is the region of the
graph which shows a linear dependence of the stress with strain. At $\Delta\sim 0.1$ a first break
down occurs as the stress drops to lower values. Note that the drop of stress is not one sudden and
continuous break, as expected from a brittle material, but rather composed of a series of small
sharp drops. Each small stress drop corresponds to opening of voids in the glass, which reduce the
stress across the system. Between the elastic zone and breakdown there is a narrow area of plastic
flow, namely more or less constant stress, while the strain increases. This area of the graph
corresponds to opening of very small holes all over the glass. However, this region is narrow, in
contrast to very long plastic flow regions that other studies have found, using  LJ~\cite{FL} and Stillinger-Weber~\cite{GPL} laws. In this respect, the AL force law is qualitatively similar to the CLJ force
of Falk.  We see also that while different in detail, the overall features of the curves for the three samples are
the same.   Also, the bulk linear elastic properties at small strain
are nearly identical, as one would expect.  Measurements on larger systems also produced very similar results.

Even though our AL force law exhibits breakdown, with only a restricted plastic flow regime, its behavior
cannot be characterized as truly brittle.  In an ideal brittle material, the stress would quickly fall to zero after
the yield stress is reached.  Here however, the stress does not fall to zero, even for strains a few times critical.
More importantly, the fall is not smooth and consists of many spikes.  This indicates the existence of many
energy barriers encountered in the fracture process.  This spikiness was studied in great detail by Gagnon,
Patton and Lacks (GPL)~\cite{GPL} (albeit in a region of plastic flow).  They showed that the origin of this
phenomenon was in the ability of the material to locally reconfigure itself into a new metastable state.  We believe
that this is at the heart of the inability of our model material to support running cracks.  The material can soak
up the elastic energy by continually finding new local metastable configurations of the crack tip region.  These
metastable configurations do not exist in a crystalline material.  This would explain why a crystal formed using the same force-law is able to support running cracks.

It is also instructive to consider the stress-strain curve for different rates of strain increase.  This is presented
in Fig. \ref{manyrates}.  We see that the general trend is that at the higher strain rates, the fall of the stress is
smoother, with fewer and smaller spikes than seen at low strain rate.  It would appear that there is a time scale
associated with the construction of a new metastable configuration.  If the strain rate is too high, the system does
not have time to find this new state.  We also verified that if the strain increase was halted, further cracking ceased
essentially immediately.

\section{Discussion}
In this work, we have seen that the model glasses we have considered do not support running
cracks. This is despite the fact that the stress-strain curve for these materials exhibits only a very small
plateau which would be indicative of ductile behavior.  
The question that remains is how to achieve brittle fracture behavior in a model of glass.  Clearly some
model must work, since experimental brittle glasses are common.  We are faced with a situation where
traditional measures insufficiently characterize the dynamical fracture behavior of a material.  A similar situation
was encountered with the standard Stillinger-Weber model of crystalline Si, which was found to be
not brittle~\cite{marderSi}, in contradiction
to the experimental situation.  A related issue is what aspect of the model is responsible for the existence/nonexistence
of running cracks.  Our guess is that it is related to the degree of crystalline order.  It seems possible to us that
if a model exhibits sufficient short-range order, it should exhibit running cracks.  This is related to the question of
time-scales discussed above in the context of the stress-strain curve.  If the crack is moving quickly enough
on the time-scale necessary to find a new metastable tip configuration, then the crack will continue to run.  It is
reasonable to suppose that the stronger the short-range order, the longer this reorganizational time-scale will
be.  We hope to test these ideas in a future work.

\end{document}